\documentstyle[aps,prb,preprint]{revtex}
\tighten
\begin{document}
\draft
\title{
$\epsilon$-Expansion of the Conductivity at the
Superconductor-Mott Insulator Transition
}
\author{Rosario Fazio$^{(1)}$ and Dario Zappal\`a$^{(2)}$}
\address{
 $^{(1)}$Institut f\"ur Theoretische Festk\"orperphysik,
Universtit\"at Karlsruhe, 76128 Karlsruhe, Germany$^{\ast}$\\
 $^{(2)}$Dipartimento di Fisica, Universit\`a di Catania, and
INFN, sezione di Catania,
corso Italia 57, 95128 Catania, Italy\\
}
\maketitle

\begin{abstract}
We study the critical behavior of the conductivity $\sigma(\omega)$ at
the  zero temperature superconductor-Mott insulator transition in
$d$  space-time dimensions for a model of bosons with
short-range interaction and no disorder.
We obtain $\sigma(\omega_n ) = (4e^2/\hbar) \sigma_{\epsilon}
\omega_n^{1-\epsilon}$, as predicted by the scaling theory, and the
prefactor $\sigma_{\epsilon}$ is calculated in the
$\epsilon$-expansion, to order $\epsilon ^2$ ($\epsilon = 4-d$).
In two spatial dimensions, ($d=3$), we find a value of  the universal
conductance $\sigma^\star =0.315 (4e^2/h)$, in  good agreement with
the known Monte Carlo results.
\end{abstract}

\pacs{PACS numbers: 74.75 +t, 74.40 +k}

\narrowtext

In two dimensions at the $T=0$ superconductor-insulator (SI) transition
the conductance is finite and universal~\cite{Fisher1,Wen}.
Since the original prediction of a metallic behavior at zero
temperature for two dimensional superconductors there has
been a lot of interest both on experimental and theoretical side.

The systems where the $T=0$ SI transition has been
observed are ultra thin superconducting films~\cite{Goldman}
and Josephson junction arrays~\cite{Geerligs,vanderZant,Chen}.
The films are driven through the SI transition by varying the
thickness while, for the arrays, the
ratio of the charging energy to the Josephson
coupling determines whether they are in the insulating or in the
superconducting phase.

Many theoretical works have investigated various aspects of the SI
transition. The superconductor-Mott insulator
transition and the universal conductance in a  model with  no
disorder was considered in Ref.~\onlinecite{Cha1} by means of  $1/N$
expansion and Monte Carlo simulations. The duality between  charges
and vortices, leading to an universal conductance of
$(4e^2/h)$~\cite{Cha1}, has been discussed in Ref.~\onlinecite{Fazio}.
Fisher {\it et al.}~\cite{Fisher2} studied the disordered
interacting boson problem using a scaling approach.
The dirty boson system and the transition to the Bose glass phase
(including the case of  long-range Coulomb interaction) was extensively
studied in~\cite{Batrouni,Sorensen}.
Wen employed a scaling theory of conserved current at anisotropic
critical points~\cite{Wen2} identifying many universal amplitudes.
One of these amplitudes in two
dimensions reduces to the universal conductance $\sigma ^{\star}$.
The finite frequencies properties close to the transition
point were analyzed by means of the $1/N$ expansion~\cite{Otterlo,Kampf}.

The determination of the value of $\sigma ^{\star}$ relies almost
entirely on numerical methods (Monte Carlo~\cite{Wallin} and
exact diagonalization~\cite{Runge}).
The only analytical calculation of $\sigma ^{\star}$  is based on the
$1/N$ expansion~\cite{Cha1}. Another powerful and widely used method
to calculate the critical properties is the $\epsilon$-expansion
(for a review see Ref.~\onlinecite{Primvan}).

In this Letter the conductivity at the SI transition is calculated
applying the $\epsilon$-expansion. We will compute the scaling form
of the $\sigma(\omega)$ discussed in Refs.~\onlinecite{Fisher1,Wen2}
and we will determine, in two spatial dimensions, the value of
$\sigma ^{\star}$.

The description of the SI transition  starts from the
hypothesis that the critical properties are determined by
the quantum  fluctuations of the phase of the superconducting order
parameter. A widely used model for the SI transition is a Hubbard like
model for strongly interacting bosons on a lattice
(Bose-Hubbard model)~\cite{Fisher2,Batrouni,Bruder}.
\begin{equation}
	H=\frac{U}{2}\sum_{i}n_{i}(n_{i}-1) -\mu\sum_{i}n_{i}
	-\frac{t}{2}\sum_{\langle ij\rangle}(b^{\dagger}_{i}b_{j}+
	h.c.)
\label{bh}
\end{equation}
where $b^{\dagger}, b$ are the creation and annihilation operators
for bosons and $n_i$ is the number of bosons.
$U$describes the short range  interaction between bosons,
$\mu$ is the chemical potential and $t$ is the hopping matrix element.
At integer fillings, the Bose-Hubbard model in  $d-1$ spatial
dimensions is in the same universality class of the XY model in
$d$ dimensions. It is more convenient for
computational purposes to work with the Ginzburg-Landau action (onto which
the Bose-Hubbard model maps)
\begin{equation}
F[\psi] =  \int d^{d-1}\vec{r} d\tau \left\{
	\delta_0 |\psi |^2 + |\nabla\psi |^2 + |\partial_{\tau}
	\psi|^2  + \frac{\kappa^\epsilon ~ u_0}{2} |\psi|^4\right\} .
\label{freeen}
\end{equation}
where $\psi(\vec{r},\tau)$ is a complex one-component scalar field.
In (\ref{freeen}) we have introduced $\epsilon=4-d$
and the arbitrary momentum scale
$\kappa$ in order to keep the quartic coupling $u_0$ dimensionless.
The bare parameters $\delta_0$ and $u_0$ can be related to the coupling
constants in the Bose-Hubbard model~\cite{Fisher2,Otterlo}.
$u_0$ includes the effect of the interaction and
$\delta_0$  measures of the distance from the critical point
(we set $\delta_0 = 0$ in the following).
It is evident from eq. (\ref{freeen}) that the quantum system at
zero temperature is equivalent to a classical system in $d$ dimensions.

The conductivity of the system can be obtained by minimally coupling
the scalar field in eq. (\ref{freeen}) to
a vector potential $\vec{A}$. The imaginary time
frequency-dependent conductivity is given by
\begin{equation}
\sigma_{\mu \nu }(\omega _n) = \frac{\hbar}{\omega _{n}}\int d^{d-1}r \, d\tau
     	   \left.
	   \frac{\delta^2\ln Z}{\delta A_{\mu}(r,\tau) \delta A_{\nu}(0)}\;
           \right|_{\vec{A}=0} e^{i\omega_{n}\tau}
\label{conductivity}
\end{equation}
where $Z$ is the partition function of the system.
The conductivity is obtained by analytically continuing
eq. (\ref{conductivity}) to real frequencies
($i\omega_n=\omega+i0^+$). Being interested in the zero field
case, we will only consider the longitudinal conductance
(e.g. $\sigma_{xx}(\omega_n)$ ), which we shall generically indicate as
$\sigma(\omega_n)$).

By performing the functional derivatives it is straightforward to
express the conductivity in terms of two and four point Green's
functions. Following Cha et al.~\cite{Cha1} it is convenient to
introduce the notation $ \sigma(\omega _n)
=(4e^2/\hbar) \rho(\omega _n)/ \omega _n$ where
\begin{eqnarray}
\rho(\omega _n) & = & 2\int \frac{d^dq}{(2 \pi)^d}\langle
	\psi^{*}_{\vec{q}} \psi_{\vec{q}}\rangle
   - 4 \int \frac{d^dq}{(2 \pi)^d}\frac{d^dp}{(2 \pi)^d}q_x p_x
	\langle \psi^{*}_{\vec{q}-\frac{\vec{k}}{2}}
        \psi^{*}_{\vec{p}+\frac{\vec{k}}{2}}
	\psi_{\vec{p}-\frac{\vec{k}}{2}} \psi_{\vec{q}+\frac{\vec{k}}{2}}
        \rangle \;.
\label{rho}
\end{eqnarray}
In eq. (\ref{rho}) the $\vec{q}$ are real
vectors in the $d$-dimensional space  and $\vec{k}=(0,0,...,\omega_n )$.

The two-point function in eq. (\ref{rho}), expressed in terms of the
self-energy
$\Sigma(\vec q)$ is (in the following we shall use the notation
$q=|\vec{q}|$):
$\langle \psi^{*}_{\vec{q}} \psi_{\vec{q}}\rangle=G(\vec q)=\left(q^2+
\Sigma(\vec q)\right)^{-1}$.
Since the three point irreducible vertex
is zero in the insulating phase where  no spontaneous symmetry breaking
occurs,
the four point function in eq. (\ref{rho}) is the sum of
a connected part containing the irreducible four point vertex
$\Gamma^{(4)}(\vec q,\vec p,\vec k)$,
and two non-connected parts of which only one gives non vanishing
contribution to $\rho(\omega _n)$.
$\Gamma^{(4)}(\vec{q},\vec{p},0)$ and $\Sigma(\vec{q})$ are related by
the following Ward identity
\begin{equation}
	\frac{\partial }{\partial q_x}\Sigma(\vec{q}) =
	2 \int \frac{d^dq}{(2 \pi)^d} p_x G^2(\vec{p})
	 \Gamma^{(4)}(\vec{q},\vec{p},0)\;.
\label{ward}
\end{equation}
This identity, related to the underlying gauge symmetry of the free
energy in eq.(\ref{freeen}),
is essential to prove that  $\rho(\omega _n =0) = 0$. This can be
easily checked by evaluating the derivative of the two-point Green's
function and substituting in eq. (\ref{rho}). In the
superfluid phase the cancelation does not occur due to the
presence of  three point vertex in the theory.

In order to get a more suitable expression of
$\rho(\omega _n)$ for a perturbative analysis, we split the four point
irreducible vertex in five parts,
$
\Gamma^{(4)}(\vec{q},\vec{p},\vec{k}) =
\Gamma^{(4)}_o+
\Gamma^{(4)}_s(\vec{q}+\vec{p}) +
\Gamma^{(4)}_t(\vec{q}-\vec{p}) +
\Gamma^{(4)}_u(\vec{k})+
\Gamma^{(4)}_{res}(\vec{q},\vec{p},\vec{k})
$
according to the external momenta dependence.
$\Gamma^{(4)}_o$ has no momentum dependence and
$\Gamma^{(4)}_{res}(\vec q,\vec p,\vec{k})$ is the residual part with
momentum dependence different from  $\vec q +\vec p$
or $\vec q -\vec p$ or $\vec k$.
This splitting turns out to be interesting because
$\Gamma^{(4)}_o$ and $\Gamma^{(4)}_u(\vec{k})$, when inserted in eq.
(\ref{rho}), vanish
and $\Gamma^{(4)}_{res}(\vec q,\vec p,\vec{k})$
gives a contribution to eq. (\ref{rho}),
in the $\epsilon$-expansion, only of order $O(\epsilon^3)$.
Eq.(\ref{ward}) and the decomposition of the $\Gamma^{(4)}$
allow, after some algebra, to express $\rho(\omega _n)$
in the form
\begin{eqnarray}
&& \frac{\rho(\omega _n)}{4} =\int \frac{d^dq}{(2 \pi)^d}
	\left[q^2_x +
	q_x \frac{\partial }{\partial q_{x}}\Sigma(\vec{q})\right]
	 G(\vec{q})
	\left[G(\vec{q}) -G(\vec{q}+\vec{k})\right] \nonumber \\
&&	- \int \frac{d^dq}{(2 \pi)^d}\frac{d^dp}{(2 \pi)^d}q_x p_x
	G(\vec{q})G(\vec{p}))
        \Biggl\{\left[G(\vec{q}+\vec{k})- G(\vec{q})\right]
	\gamma^{(4)}(\vec{q}-\vec{p})
	\left[G(\vec{p}+\vec{k})- G(\vec{p})\right]
	\nonumber \\
&&	-G(\vec{p})\left[2~G(\vec{q}+\vec{k})- G(\vec{q})\right]
        \Gamma_{res}^{(4)}(\vec{q},\vec{p},0)
        + G(\vec{p}+\vec{k})G(\vec{q}+\vec{k})
        \Gamma_{res}^{(4)}(\vec{q}+\frac{\vec{k}}{2},\vec{p}+
        \frac{\vec{k}}{2},\vec{k})\Biggr\}
\label{rho3}
\end{eqnarray}
where we have introduced
$
\gamma^{(4)}(\vec{q}-\vec{p})=\Gamma^{(4)}_t(\vec{q}-\vec{p}) -
\Gamma^{(4)}_s(\vec{q}-\vec{p})
$.
So far no approximations have been used to obtain eq. (\ref{rho3}).
The next step is the computation of the self-energy and four point
vertex in the framework of the $\epsilon$ -expansion.

At the critical point $G^{-1}(\vec{q})=q^{2-\eta}$
and the value of the critical exponent $\eta$,
for the theory considered, is known to be
$\eta=(\epsilon^2/50)(1+19\epsilon/20)+O(\epsilon^4)$.
The expansion of $G(\vec{q})$ in powers of
$\epsilon$ follows straightforwardly.

Let us now consider the four point vertex
$\Gamma^{(4)}(\vec{q},\vec{p},\vec{k})$;
its expansion in terms of the dimensionless
renormalized coupling constant $u$ can be formally written as
\begin{eqnarray}
\Gamma^{(4)}(\vec{q},\vec{p},\vec{k})=\kappa^\epsilon\left [
-2u+u^2~A(\vec{q},\vec{p},\vec{k})+
u^3~B(\vec{q},\vec{p},\vec{k})+O(u^4)\right ]
\label{renor}
\end{eqnarray}
It is understood that the function in eq. (\ref{renor}) is renormalized
and therefore, employing dimensional regularization, no pole
at $\epsilon=0$ is present in
$A(\vec{q},\vec{p},\vec{k})$ and $B(\vec{q},\vec{p},\vec{k})$.
In the following calculation the minimal subtraction renormalization
prescription has been used.

Since we are concerned with the computation of the four point vertex at the
transition, we must evaluate eq. (\ref{renor}) at $u=u^*$, $u^*$ being the
fixed point of the $\beta$ function of the theory: $\beta(u^*)=0$.
By introducing the counterterms $a_i$ through the expansion of the
bare coupling $u_0$ appearing in eq. (\ref{freeen}):
$u_0=u(1+a_1 u+a_2 u^2+...)$,
we find from the evaluation of the bare four point vertex at two loop
in the minimal subtraction scheme:
$
a_1=5/(8\pi^2\epsilon),\;
a_2=
\left (25/\epsilon^2-15/(2\epsilon)\right)/(8\pi^2)^2 \; .
$

The fixed point $u^*$, obtained as a function of the counterterms
$a_i$ (see for instance Ref. \onlinecite{amit}),  is
$
u^*=(8\pi^2)\left(\epsilon/5+3\epsilon^2/25\right)
+O(\epsilon^3)\; .
$
Since the coefficients $A(\vec{q},\vec{p},\vec{k})$ and
$B(\vec{q},\vec{p},\vec{k})$ do not contain any pole at $\epsilon=0$,
a full $O(\epsilon^2)$ calculation of the four point vertex requires
the computation of the one loop part $A(\vec{q},\vec{p},\vec{k})$ only.

Nevertheless, let us consider the momentum dependent structure
of both $A(\vec{q},\vec{p},\vec{k})$ and
$B(\vec{q},\vec{p},\vec{k})$. In the one loop part
$A(\vec{q},\vec{p},\vec{k})$ one has terms like
$\epsilon^n~log^{n+1}$,
$\epsilon^{n+1}~log^{n+1}$,
$\epsilon^{n+2}~log^{n+1}$,..., with $n=0,1,2,..$ and $log$ being the logarithm
of a generic combination of external momenta; in the two loop part
$B(\vec{q},\vec{p},\vec{k})$ one gets
$\epsilon^n~log^{n+2}$,
$\epsilon^{n}~log^{n+1}$,
$\epsilon^{n+1}~log^{n+1}$,....
As mentioned above, a full $O(\epsilon^2)$ computation of eq. (\ref{renor})
requires just the coefficient of the term
$\epsilon^0~log$ in $A(\vec{q},\vec{p},\vec{k})$, but, since we are interested
in summing formally the leading $log$ series in eq. (\ref{renor}),
we need to know at least the first two terms of the series.
This amounts to calculate the terms
$\epsilon~log^{2}$ in $A(\vec{q},\vec{p},\vec{k})$ and
$\epsilon^0~log^{2}$ in $B(\vec{q},\vec{p},\vec{k})$.

In the minimal subtraction scheme
$A(\vec{q},\vec{p},\vec{k})$ is
\begin{eqnarray}
&& A(\vec{q},\vec{p},\vec{k})=\frac{2}{(8\pi^2)}\left[ C -
       \log\frac{|\vec{q}+\vec{p}|}{\kappa}
       -2~\log\frac{|\vec{q}-\vec{p}|}{\kappa}
       -2~\log\frac{|\vec{k}|}{\kappa}\right. \nonumber \\
&&     \left. +\frac{\epsilon}{2}~\log^2\frac{|\vec{q}+\vec{p}|}{\kappa}
       +\epsilon~\log^2\frac{|\vec{q}-\vec{p}|}{\kappa}
       +\epsilon~\log^2\frac{|\vec{k}|}{\kappa}+O(\epsilon)\right]
\label{oneloop}
\end{eqnarray}
where, according to our task, we have displayed all the contributions
which are relevant
for the computation of the first two terms of the leading $log$
series and neglected the subleading logarithms
($\epsilon~log$) and the higher powers of $\epsilon$.
$C$ in eq. (\ref{oneloop}) is a $O(1)$ number,
irrelevant for the determination of
$\rho(\omega_n)$ in eq. (\ref{rho3}).

Still we need the
$\epsilon^0~log^{2}$ part of $B(\vec{q},\vec{p},\vec{k})$, which we indicate
as $\hat B(\vec{q},\vec{p},\vec{k})$; it is
\begin{eqnarray}
     \hat B(\vec{q},\vec{p},\vec{k})=
     -\frac{10}{(8\pi^2)^2}\left(
       \log^2\frac{|\vec{q}+\vec{p}|}{\kappa}
       +2~\log^2\frac{|\vec{q}-\vec{p}|}{\kappa}
       +2~\log^2\frac{|\vec{k}|}{\kappa}\right)
\label{twoloop}
\end{eqnarray}
As stated above, to order $\epsilon^2$,
$\Gamma_{res}^{(4)}(\vec{q},\vec{p},\vec{k})=0$, making eq. (\ref{rho3})
much simpler. Actually the first nonvanishing contribution
to  $\Gamma_{res}^{(4)}(\vec{q},\vec{p},\vec{k})$ comes from the
subleading term $\epsilon^0~log$ in $B(\vec{q},\vec{p},\vec{k})$
which is neglected in our calculation.
We are now able to sum formally the leading $log$ series obtaining some power
of the momenta involved (note that the momentum-independent term
$O(\epsilon)$ in the series is irrelevant because it vanishes when
integrated in eq. (\ref{renor})).
Putting together eqs.(\ref{renor}, \ref{oneloop},
\ref{twoloop}), $\gamma^{(4)}(\vec{q}-\vec{p})$ becomes
\begin{eqnarray}
\gamma^{(4)}(\vec{q}-\vec{p})=
-\frac{16\pi^2\epsilon}{25}|\vec{q}-\vec{p}|^\epsilon+O(\epsilon^3)
\label{gamf}
\end{eqnarray}
The importance of summing the leading $log$ series is now clear:
eq.  (\ref{gamf}) is correct only to order $O(\epsilon^2)$, but the
infrared singularities due to the logarithms in eqs.
(\ref{oneloop}, \ref{twoloop}), disappear in the
resummed quantity $|\vec{q}-\vec{p}|^\epsilon$.

Another comment is in order: the last result
does not depend on the arbitrary momentum scale $\kappa$
introduced in eq. (\ref{freeen}), nor on the
choice of the minimal subtraction scheme.
Indeed, one can easily prove that
the shift of the counterterms $a_i$ by a finite (no poles at $\epsilon=0$)
amount, $a_i\to a_i+\overline{a_i}$, does not change
$\gamma^{(4)}(\vec{q}-\vec{p})$ in eq. (\ref{gamf}).

Finally, going back to eq. (\ref{rho3}), we can insert eq. (\ref{gamf}),
and discard the terms containing
$\Gamma_{res}^{(4)}(\vec{q},\vec{p},\vec{k})$, thus obtaining
$\rho(\omega _n)$ and, consequently, $\sigma(\omega _n)$ to order $\epsilon^2$;
in dimensions lower than $4$
all integrals in eq. (\ref{rho3}) are convergent in the
ultraviolet (in $d \ge 4$, a short-range cutoff should be introduced
to regularize the results, but this is not important for our purposes)
and we find
\begin{eqnarray}
&& \sigma(\omega_n) = \frac{4e^2}{\hbar}
\left\{ \left( \frac{\epsilon^2}{25} - 2   \right)
\frac{ \Gamma^2(d/2) \Gamma(1-d/2)}{\pi^{d/2}~2^d~\Gamma(d)}
	- 2~z(d) \int \frac{d^d~y}{y^{4}}
        f_{\vec{k}} (\vec{y}) f^*_{\vec{k}}(\vec{y})
            \right\} \omega_n^{1-\epsilon}
\label{central}
\end{eqnarray}
where
\begin{eqnarray}
f_{\vec{k}}(\vec{y})=\int_0^1 d\alpha\int_0^\infty
\frac{d\lambda}{(4\pi)^{d/2}}~\frac{y_x}{2~\lambda^{d/2}}~
exp \left( -\frac{y^2}{4\lambda} \right)
\Biggl[ 1 - exp\Bigl(-\alpha(1-\alpha)\lambda-i\alpha y_z  \Bigr)   \Biggr]
\label{effe}
\end{eqnarray}
and
$
z(d) = \left[\pi^{d/2}/(\Gamma(d/2))
            \int_0^\infty dx~x^{d-5}
            \left(  1- \left(  x/2  \right)^{1-d/2}
             \Gamma \left( d/2 \right) ~J_{(d/2-1)}(x)   \right)
            \right]^{-1}
$.
$\Gamma(x)$ and $J_\nu(x)$ are, respectively, the Gamma and Bessel
functions
and the $\omega_n$ independent factor in eq. (\ref{central})
has been expressed in integral form
for the sake of simplicity.

Eq.(\ref{central}) is the central result of our paper.
The  form of the frequency dependent conductivity, as  predicted from
the scaling theory~\cite{Fisher1,Wen2}, is obtained.

In the calculation of the $\sigma(\omega _n)$, $\epsilon$-expansion
was used to determine the irreducible vertices. The $d$-dimensional
integrals in eq.(\ref{rho3}) have been instead performed not resorting to
any expansion. This observation is crucial in
order not to obtain spurious dependences on the ultraviolet cut-off
needed  to regularize the theory in higher dimensions. In this sense
the calculation of the conductivity to O$(\epsilon ^2)$ should be
understood as the scheme of approximation for the vertices.

In two spatial dimension the universal conductance
$\sigma^\star$
can be  evaluated from eq. (\ref{central}) at $d=3$;
this leads to the result displayed in Table 1, together with the other
known determinations of $\sigma^\star$.
Our estimate is in agreement with the results of the Monte Carlo
simulations in Ref.~\onlinecite{Cha1} within $10\%$ but it should be noted
that the $1/N$ and $\epsilon$-expansion approach the Monte Carlo value
from different sides.

As a final remark we want to go back to the connection with the
experiments. The model with only phase fluctuations is certainly applicable
to Josephson junction arrays, but it has been recently questioned
in the case of thin films.
New measurements~\cite{Yazdani} point out
that the value of the conductance may be not universal and this should
be associated to the relevance of fermionic degrees of freedom at the
transition. This is also the outcome of recent calculations which
include fermions starting from a Bose model with local
damping~\cite{Wagenblast} or simulating a model of
electrons interacting through
an attractive potential \cite{Trivedi}.

\acknowledgments
We would like to thank A. Bonanno, V. Branchina, C. Bruder, A. van Otterlo,
G. Sch\"on, K-H. Wagenblast, and G.T. Zimanyi
for useful discussions. The financial support of the European
Community (HCM-network CHRX-CT93-0136),
the Deutsche Forschungsgemeinschaft
through SFB 195 and INFN  is gratefully acknowledged.

\begin{table}
$$
\begin{array}{c c c c}
\hline \hline
\mbox{Method} & \epsilon \mbox{-exp}^{(a)} &
\displaystyle{(1/N)\mbox{-exp}^{(b)}}
& M C^{(c)} \vspace{0.1cm} \\ \hline
 &          &                            & \\
\hspace{0.5cm} \displaystyle{\frac{h}{4e^2}\sigma^{\star}}= \hspace{0.5cm} &
\hspace{0.5cm} 0.315
\hspace{0.5cm} & \hspace{0.5cm} 0.251
\hspace{0.5cm}
& \hspace{0.2cm} 0.285 \hspace{0.2cm} \vspace{0.1cm} \\ \hline \hline
\end{array}
$$
\caption{The value of the universal conductance obtained by various
methods are reported for comparison. a) this work, b) - c)
Ref.\protect\onlinecite{Cha1}}
\end{table}

\end{document}